\documentclass[twocolumn,showpacs,preprintnumbers,amsmath,amssymb]{revtex4}

\usepackage{graphicx}% Include figure files
\usepackage{dcolumn}% Align table columns on decimal point
\usepackage{bm}% bold math

\begin{document}

\preprint{APS/123-QED}

\title{Topology and Evolution of Technology Innovation Networks}

\author{Sergi Valverde$^1$, Ricard V. Sol\'e$^{1,2}$, Mark A. Bedau$^{3,4}$ and Norman Packard$^{2,4}$}
% \homepage{http://complex.upf.es/~sergi}
\affiliation{
$^1$ ICREA-Complex Systems Lab, Universitat Pompeu Fabra, Dr. Aiguader 80, 08003 Barcelona, Spain\\
$^2$Santa Fe Institute, 1399 Hyde Park Road, New Mexico 87501, USA \\
$^3$Reed College, 3203 SE Woodstock Blvd., Portland 97202, USA\\
$^4$ProtoLife, Parco Vega, Via della Liberta 12, Marghera 30175, Venice, Italy}


\begin{abstract}
The web of relations linking technological innovation can be fairly described in terms of patent
citations. The resulting patent citation network provides a picture of the large-scale organization of
innovations and its time evolution. Here we study the patterns of change of patents registered by the
US Patent and Trademark Office (USPTO). We show that the scaling behavior exhibited by this
network is consistent with a preferential attachment mechanism together with a Weibull-shaped
aging term. Such attachment kernel is shared by scientific citation networks, thus indicating an
universal type of mechanism linking ideas and designs and their evolution. The implications for
evolutionary theory of innovation are discussed.
\end{abstract}

 % Networks and genealogical trees
 % Fluctuation Phenomena, random processes, noise and Brownian motion
 % Dynamics of Evolution

%Use showkeys class option if keyword

\maketitle

%---------------------------------------------------------------------------

\section{Introduction}

%---------------------------------------------------------------------------    

Innovation takes place both in nature and technology \cite{Erwin}. Either 
through symbiosis \cite{Margulis}, tinkering \cite{Jacob} or design \cite{Complex,Lienhard} new functional structures 
and artifacts are obtained. Such new entities often result from the combination 
of predefined designs or building blocks, although a completely new solution 
can also emerge. This is the case for example of the replacement of vacuum tube
technology by semiconductors. However, the majority of technological (and 
evolutionary) changes take place by means of a progressive path of change. Such
steady and successful transformation of designs is largely based on an extensive 
combination and refinement of existing designs.

A surrogate of the ways in which innovations take place in time is provided 
by patent files. Patents are well-defined objects introducing a novel design,
method or solution for a given problem or set of problems. Additionally, they 
indicate what previous novelties have been required to build the new one.
In order to gain insight into the global organization of the
patterns of innovation and their evolution in technology, here we study a very
large data base including all USPTO patents from 1975 to 2005 \cite{USPTO}.

\begin{figure}[htbp]
\begin{center}
\includegraphics[width=.5\textwidth]{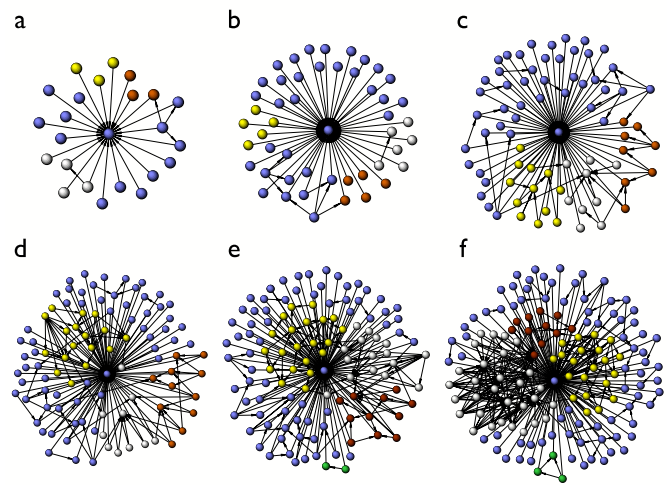}
\caption{ From (a) to (f), evolution of a patent subset related to computed tomography. 
The hub in the center corresponds to the precursor invention by G. Hounsfield 
(US patent 3778614).} 
\label{fig_tomoevo}
\end{center}
\end{figure}  

As it occurs with the fossil record for evolution, the record of patents through
time provides us with the opportunity of seeing how new inventions emerge and 
how they relate to previous ones. A given patent will typically require 
new solutions and previously achieved results. Looking at how patents link to 
each other is the simplest way of having a large scale picture of the patterns 
and processes associated to the collective dynamics of innovation unfolds \cite{book,Kuznets}. 
Many interesting questions can be formulated in relation to this: what is the
global organization of interactions among innovations? Is this a repeatable
pattern? How are similar classes of innovations related among them? Do
these patterns respond to history-dependent rules or are instead describable
by means of simple models? These questions are addressed here and it is
shown that a standard statistical physics approach provides a good picture
of how these webs emerge.

The paper is organized as follows: in section II the data set analysed is presented.
In section III the topological trends exhibited by the patent citation network are discussed under
the light of a model of graph growth with aging (section IV). In section
V our basic results are summarized and its implications outlined.

%---------------------------------------------------------------------------

\section{Patent Citation Networks}

%---------------------------------------------------------------------------

Previous studies have measured the value of an innovation by means of the analysis of patent 
citations, i.e., the rate of receiving new citations. However, innovation is an elusive 
notion that is difficult to measure properly and existing measures provide limited insight 
 \cite{book}. It is a difficult task to find useful indicators for the
value of innovations. In this context,  we introduce patent citation networks 
as an appropriate approach to the global analysis of
the process of technological innovation.  Recent work in complex networks provides several models 
that describe or reproduce structural features of natural and artificial evolving systems.  Here,
we will show how innovation can be described as a process of network growth following some specific 
rules. In particular, our model provides a rigorous statistical test to assess the balance between
 patent importance and patent age, i.e., Price's "immediacy factor" \cite{book}. 

The set of patents and their citations describes a (so-called) patent citation network $G$.  The
patent network belongs to the general class of citation networks, which includes the scientific citation
network. Here, nodes $v_i \in G$ represent  individual patents and the directed link $(v_i, v_j)$ 
indicates that patent $v_i$ is a descendent of patent $v_j$.  In order to illustrate the power of the
 network approach, we have re-analyzed the evolution of a well-know patent dataset. Figure 1
  shows the time evolution for the subset of patents in Computer Tomography (CT), from 1973 to
  2004.  A smaller subset of this dataset was analysed in \cite{Tomo}.  The figure indicates that
  some patents receive much more citations than others. In particular, the hub at the center
  corresponds to the very first patent in CT associated with its invention by G. Hounsfield.  

Interestingly, the network analysis reveals some other patterns that cannot be easily 
recovered by other means. For instance, in figure 1 we can appreciate the modular 
organization of the CT patents.  Here we have used Clauset et al. algorithm \cite{Clauset} to detect
community structure in large networks. Roughly speaking, topological modules are defined as
groups of nodes having more links among them than with other elements in the graph. Thus, 
patents belonging to the same module share a common color. Although we have not 
explored this problem in detail, direct inspection of the networks shown in figure 1 reveals
that the modular structure seems to correlate well with shared functional traits. As an example,
the white module involves several related patents associated with X-ray tomography.

\begin{figure}[htbp]
\begin{center}
\includegraphics[width=.43\textwidth]{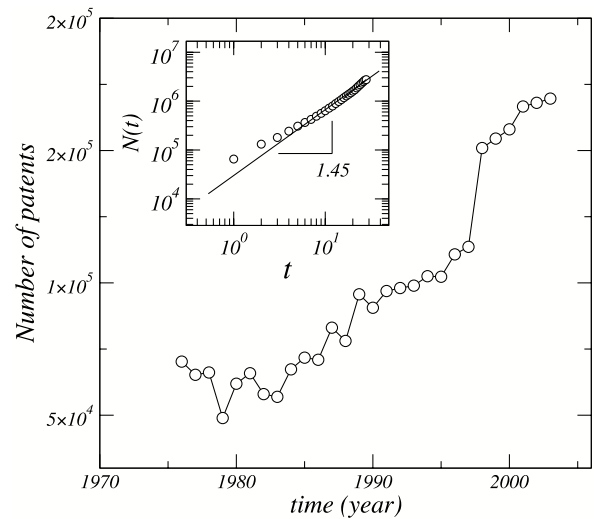}
\caption{ Time evolution of the number of patents $N(t)$ in the USPTO dataset from 1973 to 2004. 
 Inset: Cumulative number of patents on a log-log scale, showing a scaling $N(t) \sim t^{\theta}$. } 
\label{fig_size}
\end{center}
\end{figure}  

Beyond specific patterns of patent evolution, here we aim to detect universal trends in the
global evolution of the US patent system.   The patent citation network (PCN) analyzed here has 
$N=2801167$ nodes and  $L=18053661$ links. Its time evolution from 1976 to 2005 is shown
 in figure \ref{fig_size}. The number
 of patents at a given time $t$ scales as a  power law:

\begin{equation}
N(t) \sim t^{\theta}
\label{growth}
\end{equation}
with an exponent $\theta = 1.45 \pm 0.06$. Some recent papers have explored the 
patent citation datasets at different levels, including a graph theoretical approach on 
a large scale \cite{Csardi} or involving a more specific case study, such as fuel cell 
research \cite{Verspagen}. Here we will show that the statistical 
features of this network can be explained by 
using an appropriate attachment kernel describing how successful patents 
become more linked and how this preferential attachment decays with age.

%---------------------------------------------------------------------------

\section{Distribution of Patent Citations}

%---------------------------------------------------------------------------

Citations are often interpreted as indicators of innovation size or economic value
\cite{Trajtenberg}. The distribution of innovation size (defined as the number of citations to a patent)
 is skewed \cite{ Kuznets, Scherer65, Scherer98}. However, there is an ongoing discussion about 
 the particular nature of this distribution. In particular,
there is no general agreement whether it follows a log-normal or Pareto distribution \cite{Harhoff, Scherer98} .  
Still, there are common patterns like the existence of some extreme values, which is consistent with
 a power-law tail.  We report similar features in the in-degree distribution studied here (see below).

\begin{figure}[htbp]
\begin{center}
\includegraphics[width=.45\textwidth]{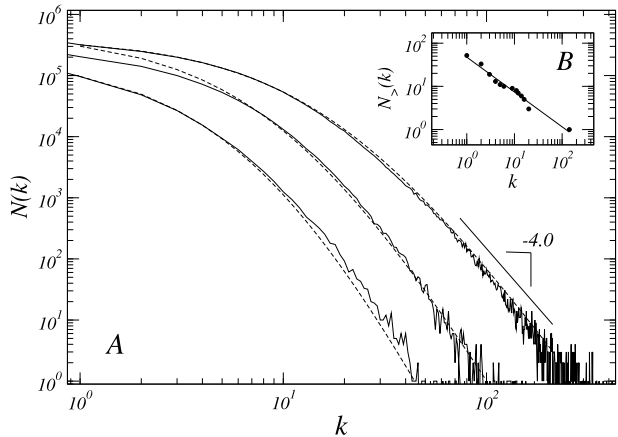}
\caption{ (A) The in-degree distribution for the patent citation network follows an extended
power-law distribution, $P_i(k) \sim (k + k_0)^{-\gamma}$.  Three distributions are 
displayed for three different time windows, namely 1984 (leftmost), 1992 (center) and
 2002 (rightmost). (B) The in-degree distribution for the subset of patents displayed in 
fig.1 f (for computer tomography) is roughly approximated by a scale free distribution. 
The leftmost point indicates the central hub in fig.1. } 
\label{fig_indegree}
\end{center}
\end{figure}

The in-degree distribution $P_i(k)$ is equivalent to the so-called distribution of number 
of patent citations. Figure \ref{fig_indegree}A shows the in-degree distribution for the 
patent citation network in 2004. Notice that $P_i(k)$ is neither exponential nor a simple power law. 
Instead, we have found that an extended power-law form fits the in-degree distribution very well:

\begin{equation}
P_i(k) \sim (k + k_0)^{-\gamma}
\label{indeg}
\end{equation}

where $k_0 = 19.46 \pm 0.22$ and $\gamma=4.55 \pm 0.04$.  This extended power-law
reduces to a power-law when $k \gg k_0$ and it degenerates to an exponential distribution 
for $k \ll k_0$. The extended power-law distribution has been related to a mixed attachment
mechanism \cite{Zhou}. However, here we will show that this explanation does not apply 
for the patent citation network.  Instead, we propose that the extended power-law form for
the in-degree distribution stems from a combination of both preferential attachement and aging 
\cite{DorogovtsevMendes2000}.

%---------------------------------------------------------------------------

\section{Evolution}

%---------------------------------------------------------------------------

Lets us assume that every patent has a unique identifier $0 < i < t$. Our model starts at time $t=0$ 
when there is only one patent in the network. From this initial network, we add a new patent $i$
at every time step that references $m$ previous patents.  
Two main forces drive the evolution of the patent citation network.  First, it is natural to 
assume that the number of patent citations (i.e., incoming links) is a surrogate of its relevance
\cite{PatentBook}. Useful patents are more likely to receive further citations than marginal
inventions. Thus, the probability of receiving new citations should be proportional to
the current number of citations. This rule parallels the preferential attachment mechanism
of network growth \cite{BA}. Under this rule new elements entering the system
connect with other nodes with a probability $\Pi (k)$ that is proportional to its 
degree, i.e., 

\begin{equation}
{\Pi (k) \sim k}
\label{pref}
\end{equation}
However, old patents tend to be less relevant in the context of recent innovations:
 attachment rates decay as the patent losses value. In particular, 
 patents are released to the public domain after some finite period of exploitation.
 
 The evolution of complex networks involving both preferential attachment and aging has
 been extensively studied. In particular, Dorogovtsev and Mendes (DM)
 determined analytically the scaling properties of the resulting networks  \cite{DorogovtsevMendes2000}.
 In the DM model, the rule of attachment scales now as:
 
 \begin{equation}
 \Pi (k, \tau) \sim k {\tau}^{-\alpha}
 \end{equation}
 where $\tau = t - i$ indicates the age of the $i-th$ node and the exponent $\alpha$ (which is positive) weights how fast is the aging affecting the likelihood of attachment. Extensions of this attachment probability kernel include accelerated growth with $ \Pi (k, \tau) \sim k^\beta {\tau}^{-\alpha}$  and 
exponential aging kernel $ \Pi (k, \tau) \sim k exp(-\tau^\alpha)$ \cite{Zhou}. 

Finally, some models of scientific citation networks take into account the simultaneous
 evolution of author and paper networks \cite{Borner2004}. In these models, the rule of
  attachment behaves as:
 
\begin{equation}
 \Pi (k_i, \tau) \sim k_{i}^{\beta}{\tau}^{\alpha-1}e^{-\left( \frac{\tau}{\tau_0} \right)^{\alpha}}
 \label{eqkernel}
\end{equation}
when the time-dependent component follows a Weibull form. Here, $\tau_0$ controls the rightward 
extension of the Weibull curve. As $\tau_0$ increases, so does the probability
of citing older papers.On the other hand, small values of $\tau_0$ indicate strong aging that favors recently
published patents  \cite{Borner2004}. Here we choose the simplest assumption 
(preferential attachment $\beta=1$) and consider the aging function in eq. \ref{eqkernel}.
Consequently, the average connectivity of the $i-th$ patent at the time $t$ evolves according to the 
following equation:
  
\begin{equation}
{{\partial\overline{k}(i,t)}\over{\partial t}}={{m\overline{k}(i,t)f(t-i)}\over{\int_{0}^{t}\overline{k}(u,t)f(t-u)du}}
\label{evot}
\end{equation}
where $m$ is the number of links introduced at each step ($m=1$ is the DM model).  Now we
address the following question: is the above equation consistent with the patent network evolution?  
 In the following, we will estimate the form of  the  attachment kernel (and the corresponding $\alpha$, $\beta$
  and $\tau_0$ parameters) for the patent citation data.  
  
First,  we consider system size $N$ as our time index instead of real time $t$. This way we avoid any bias 
 due to the pattern of non-linear growth (\ref{growth} and attach to the standard formulation of network
 models. Then, eq. (\ref{evot}) becomes:
   
\begin{equation}
{{\partial\overline{k}(i,N)}\over{\partial N}}={{m\overline{k}(i,N)f(N-i)}\over{\int_{0}^{t}\overline{k}(u,t)f(N-u)du}}
\label{evon}
\end{equation}
   
Using  ${\partial\overline{k}}/{\partial N} =  ({\partial\overline{k}}/{\partial t})  ({\partial t}/{\partial N})$ and
 the time-dependent scaling $N(t) = At^\theta$, we have:

\begin{equation}
{{\partial\overline{k}(i,N)}\over{\partial N}}= 
 \left({1 \over \theta} N(t)- A)^{{1 \over \theta}-1} \right)
 {{\partial\overline{k}(i, t)} \over {\partial t}}\end{equation}

 Now the whole time interval $N$ is partitioned into $N/ \Delta N$ time slots comprising the same number 
  $\Delta N \ll N$ of patents.  Here, $N \approx 2.8$ million patents corresponding to the time interval 
  1976-2005.  The $s-th$ time slot has the same number of  new  $\Delta N= 10^5$ patents.

\begin{figure}[htbp]
\begin{center}
\includegraphics[width=.5\textwidth]{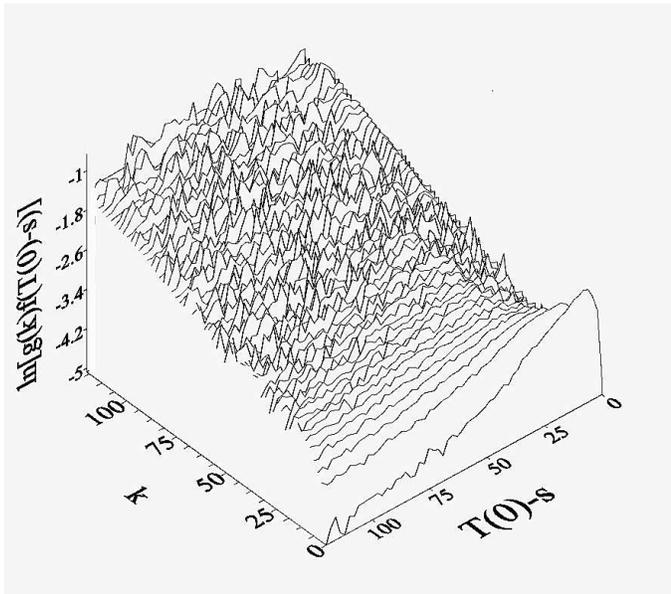}

\caption{The normalized attachment kernel $\Pi (k, \tau) \sim g(k)f(\tau)$ determined numerically for the patent
citation network at $T_0=2002$.} 
\label{fig_kernel}
\end{center}
\end{figure} 

To measure the attachment rule $\Pi (k_i, \tau)$ we monitor to which old patent new patents link, as a 
function of in-degree $k_i$ and age $\tau$ (recall that here $\tau$ is measured in number of time slots). 
We have assumed that attachment is the product of a preferential attachment function $g(k)$ and an 
aging function $f(\tau)$:

\begin{equation}
\Pi (k, \tau)  \sim g(k) f (\tau)
\label{akernel}
\end{equation} 

Following \cite{measure}, we study the citation process in a relatively short time frame (a time slot $\Delta N$).
The large number of nodes in the system (in the order of $10^6$ nodes) ensures that we will gather sufficient
 samples to recover the attachment kernel. We divide the evolution of the system in three stages: (i)
  the system before slot $T_0$, (ii) the system between slots $T_0$ and $T_1= T_0 + 1$ and
(iii) the system after $T_1$. 
 When a $T_1$ node joins the system we record the age $\tau$ and the in-degree $k$ of the
   $T_0$ node to which the new node links.  We count all the citations made by new nodes
   between $T_1$ and  $T_1+ 1$. The number of citations received by nodes $T_0$ from $T_1$ 
   nodes normalized by the in-degree frequency $P(k)$ is an approximation to the attachment
    kernel (see fig. \ref{fig_kernel}).

\begin{figure}[htbp]
\begin{center}
\includegraphics[width=.40\textwidth]{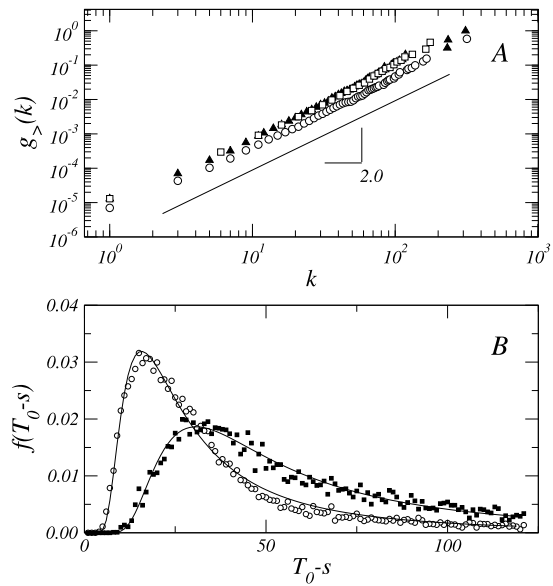}
\caption{Estimation of the attachment rule for the patent citation network at $T_1= 2003$. (a) 
Preferential attachment function fits a scaling-law $g_>(k) \sim k^{\beta+1}$ with  $\beta = 1$.  Each 
curve corresponds to nodes having the same age.  (b) Fitting for the aging function $f(\tau)$ 
predicts the Weibull distribution described in the text with $\alpha \approx -1.45$. Each 
curve corresponds to nodes having the same in-degree ($k=1$ for white balls and $k=5$ for
solid balls) . For every curve we have used $T_0 = T_1-1$. } 
\label{fig_aging}
\end{center}
\end{figure} 
 
 Using our dataset, we have estimated that $g(k) \sim k^\beta$ and found $\beta =1$, which further validates 
 our assumption of preferential attachment (see fig. \ref{fig_aging}A). Notice that in our fittings we have 
 used the cumulative function $g_>(k)= \int_{0}^{k}{g(k)dk}$ to reduce the noise level. On the other hand, 
 fig. \ref{fig_aging}B shows the Weibull distribution, which fits very well the aging function $f(\tau)$:

\begin{equation}
f(\tau) \sim \alpha \tau_0^{-\alpha} {\tau}^{\alpha-1}e^{-\left( \frac{\tau}{\tau_0} \right)^{\alpha}}
\end{equation}
with an exponent  $\alpha \approx -1.45$ and  $\tau_0 \approx 40$. An obvious advantage of using the 
Weibull form is that in naturally includes as limit cases both exponential and Gaussian 
distributions.

The common structure of the aging term found here and in the network of paper
citations  \cite{Borner2004} suggests that common patterns of organization and evolution might 
be shared. The paper citation graph, obtained by looking at the list of references included in 
each paper, is in fact close to the basic rules defining the patent citation graph. 
In both cases, cross-links are associated to some underlying set of features which 
are shared by patents or papers. As it occurs with the patent case, new papers
are based on previous ones providing the background required to build a new idea. On the
other hand, as new ideas and concepts develop into well-defined areas, they will tend to 
attach less to more generic or older works. Additionally, the observed modular
organization which might also contribute to deviate from the simple power-law attachment assumed in
previous theoretical studies. What seems clear is that there might be some universal trends 
canalizing the growth of innovation networks, whether scientific or technologic.

% ---------------------------------------------------------------------------

\section{Discussion}

%---------------------------------------------------------------------------

The patterns of innovation emerging in our society are the outcome 
of an extensive exchange of shared information linked with the 
capacity of inventors to combine and improve previous designs. Even 
very original inventions are not isolated from previous achievements. 
A patent can be identified as an object which needs a minimum amount 
of originality to be considered as truly different from previous 
patents. Moreover, to be obtained, it must properly refer to related 
patents in a fair way. Such constraints make this system specially 
interesting since we can wisely conjecture that it represents the 
expansion of real designs through some underlying technology landscape. 
These designs can be just small improvements or large advances. Our analysis 
provides a quantitative approach to this evolving structure using the 
approach of statistical physics.

We have shown that the underlying rules of network change for our 
system reveal a mixture of preferential attachment favouring 
a rich-gets-richer mechanism together with an aging term weighting the 
likelihood of citing old patents. As the network grows, recent 
patents will tend to cite recent designs (since innovation is likely to 
involve redefining recent inventions) and be less likely to link to old 
patents. The consequence of this, as predicted by previous mean field 
models (refs) is that the expected scaling law in the degree distribution associated to 
preferential attachment kernels will be modified in 
significant ways. Here we have shown that the network of patents, defined 
by using the indegree as a surrogate of patent relevance, scales as 
$P(k) \sim (k+k_0)^{-\gamma}$ with $\gamma > 4$. This is not far from 
previous predicted scaling laws (DM) associated to preferential attachment 
and power law aging (i. e. $f(t)\sim t^{-\alpha}$ which predict 
$P(k) \sim k^{-\gamma(\alpha)}$ (with $\gamma \sim 4$ for $\alpha \sim 0.5$). 
However, the humped shape of our aging term (as described by the Weibull 
distribution) makes necessary to modify these approximations. 

As a final point in our discussion, it is worth noting that we 
have strong correlations among patents indicating a complex 
organization in modules. As shown by the example in figure 1, together 
with the nonlinearities associated with the attachment rules, 
there is some underlying community structure in the patent network 
that deserves further exploration. The emergence of modules is a natural consequence
 of the specialized features shared by related patents. 
But it might also reveal the structure of the innovation landscape itself: 
new patents related to previous ones can also be understood as improved 
solutions that explore the neighborhood of previous solutions. This view 
would provide a quantitative picture of the topology of technology 
landscapes \cite{Kauffman1,Kauffman2}. Such 
an evolutionary interpretation in terms of fitness functions will be 
explored elsewhere.

\begin{acknowledgments}

We thank Vincent Anton and Marti Rosas-Casals for useful discussions. This work has been 
supported by grants FIS2004-05422, by the EU within the 6th Framework Program 
under contract 001907 (DELIS), by the James McDonnell Foundation and by the Santa Fe Institute.

\end{acknowledgments}

\bibliography{pre} % Produces the bibliography via BibTeX.

\end{document}